\documentclass[aps,twocolumn,showpacs,showkeys]{revtex4}
\usepackage{graphicx}
\begin{document}



\title{Distinguishing the opponents in the prisoner dilemma in well-mixed populations}

\author{Lucas Wardil$^{\dag\ast}$ and
Jafferson K. L. da Silva$^{\dag}$}
\affiliation{$^\dag$Departamento de F\'\i sica, Universidade Federal de Minas Gerais, \\
Caixa Postal 702, CEP 30161-970, Belo Horizonte - MG, Brazil\\}
\date{\today}

\pacs{87.23.-n, 89.65.-s, 02.50.Le}


\begin{abstract}
Here we study the effects of  adopting  different strategies against different opponent instead of adopting the same strategy against all of them in the prisoner dilemma structured in well-mixed populations. We consider an evolutionary process in which strategies that provide reproductive success are imitated and players replace one of  their worst interactions by the new one. We set individuals in a well-mixed population so that network reciprocity effect is excluded and we analyze both synchronous and asynchronous updates. As a consequence of the replacement rule, we show that mutual cooperation is never destroyed and the initial fraction of mutual cooperation is a lower bound for the level of cooperation. We show by simulation and mean-field analysis that for synchronous update cooperation dominates while for asynchronous update only cooperations associated to the initial  mutual cooperations  are maintained. As a side effect of the replacement rule, an ``implicit punishment'' mechanism comes up in a way that exploitations are always neutralized providing evolutionary stability for cooperation.
\end{abstract}

\maketitle
 
\begin{center} 
 \textbf{1. Introduction}
\end{center}

\vspace{0.5cm} 
 
Cooperative dilemma was initially studied in the framework of classical game theory. Usually individuals have two strategies: cooperation and defection. A cooperator provides a benefit to the opponent and pays a cost for that. A defector receives the benefits if the opponent is a cooperator. This defines a material payoff. If individuals maximize their material payoff, it is well know that defection will dominate \cite{r0}. Departing from these initial ideas, evolutionary game theory has emerged and  strategy evolution in populations was studied. In this approach it is implicit assumed the principle of natural selection, where the payoff is equated to fitness and the fittest strategy survives \cite{r1}.  In this context it was shown that the classical theory is recovered in the replicator equation, where population is considered to be well-mixed, that is, a population where everybody interacts with everybody \cite{r1}.

Cooperation cannot be supported without extra mechanisms \cite{r23}. Essentially two actions take place for cooperation survival: maintenance of mutual cooperation and prevention from exploitation \cite{r3}. Cooperators can be better off only if they meet each other so that their profits exceed defectors' profits. If the individuals perceive that it is important what the opponents are doing in order to attend these two essential actions, reciprocal preferences can come up \cite{r10,r11,r12,r13}. Reciprocity means that what an individual do depends on what others do to him/her directly or indirectly. Direct reciprocity means that I choose what to do against you depending on what you do to me. Indirect reciprocity means that my behavior toward you also depends on what you do to others. Another subtle way of reciprocity is network reciprocity. Individuals are set on the vertices of a network and interact only with their neighbors. In this context, cooperators form clusters of mutual cooperation and this mutualism is viewed as reciprocity \cite{r19,r20,r21,r22,r27,r32}. But human behavior is not so simple: individuals can adopt reciprocal strategies but, motivated by internal emotion, like anger against exploitation \cite{r6}, they can punish defectors \cite{r6,r8,r9}. This would not be so intriguing, as it is just another way of reciprocal motives. But the important feature is that  individuals usually input costs to defectors at their own expenses. This behavior is called altruistic punishment, because individuals pay a cost to punish even if they never met the punished opponent again and because the punishment acts weaken the defectors and the entire population  gets better off \cite{r6}. Reputation, rewards or repeated interaction, as internal motives, they all interact with punishment motives \cite{r10,r11}. Punishment involves some subtle questions and gives rise to another evolutionary puzzle: altruistic punishment, although seemingly usual, may be a maladaptive trait as the punishers get worst payoffs \cite{r13}. 

Recently it was introduced the possibility of an ``implicit punishment'' without turn to extra individual preferences except the desire to maximize the own gain. This was accomplished by the adoption of different strategies against different opponents in the context of network reciprocity with synchronous update \cite{r15}. Instead of playing the same strategy  against all of the neighbors, individuals can choose one different strategy against each opponent. If each player strategically updates their strategies possibly imitating a successful random  neighbor  and replaces the interaction that gives the worst payoff by the imitated strategy, it was shown for square lattices that cooperation was strongly supported, even for huge defection tendency, and was robust against misjudgments of the worst interaction. The possibility of opponent differentiation introduces a mechanism of punishing without costs and without any kind of internal preferences except the desire of maximize own payoff. We call this punishment ``implicit punishment''. But in that work \cite{r15}, the possibility of adoption of different strategies was introduced in the context of network reciprocity. What happens if network reciprocity is excluded?  Here we  analyze this model in well  mixed populations,  what means that we are excluding network reciprocity effects. The other important feature of the model is the synchronous update assumption. It is well known that results may be striking different if asynchronous update is used \cite{r29,r28}. Here we analyze the model with both synchronous and asynchronous updates. We show that cooperation still remains alive, although for asynchronous update it achieves its lower bound level. We analyze the model using computer simulations and a mean-field technique.

\begin{center}
  \textbf{2. The model}
\end{center}

Let us state the model formally. We study the prisoner dilemma in a  population of size $N$  as the scenario for the cooperation problem. We consider a well-mixed population, what means that each player can interact with everybody. The strategy vector of an individual is $\vec{S}=(S_1,\ldots,S_j,\ldots,S_{N-1})$, where $S_j$ can be C (cooperation) or D (defection). So  individuals are merged in $N-1$ interactions. If in one of these interactions an individual plays C against an opponent who is playing D, we denote this interaction as (C,D) (the first entry is the strategy of the focal player and the second entry is the opponent strategy).  The payoff of a D strategy against a C strategy is $P(D,C)=b$, where $b>1$ is the defection tendency. Using the same notation, we have that $P(D,D)=\epsilon$, with $\epsilon << 1$, $P(C,C)=1$ and $P(C,D)=0$. For synchronous update, each player interacts with the other $N-1$ players, plays a round of one game against each opponent, and earns a cumulative payoff. After that, each player chooses randomly one neighbor and compare their cumulative payoffs. If the opponent cumulative payoff is bigger than its own one, it imitates the strategy the opponent is using against him/her with probability proportional to the difference of cumulative payoffs, $\Pi=|\Delta P_{cum}|/((N-1)b)$ \cite{r18}. On the other hand, if the opponent cumulative payoff is lower than its own one, the focal player remains with the same strategy. If imitation takes place, the new strategy replaces the strategy used in the interaction that gives the worst payoff. The worst payoff of the  focal player is given by the interaction  (C,D), followed by (D,D), (C,C) and (D,C). For asynchronous update, a random individual is chosen so that it can imitate and possibly replace one of its strategies like in the synchronous case. After this individual update, the entire population play a round of the game, and each player earns new cumulative payoffs, and another random individual is chosen to update its strategies. A time step consists of $N$ of such processes.

\begin{center}
  \textbf{ 3. Evolutionary analysis}
\end{center}

We made our simulations using networks of size $N=40$ and $N=100$ and evaluate the fraction of cooperation ($f_c$) adopted by all of the players in all of their interactions. If $n_c$ is  the quantity of C strategies used in all of the interactions by all of the players, we have $0\le n_c \le N(N-1)$ and $f_c=n_c/N(N-1)$. The random initial configuration consists of $50\%$ of cooperation and the averages are made over $100$ different initial conditions. We use $\epsilon =0.001$ and we show here only the case $b=2$, although we simulated the model also for other values of $b$. In fact, the $b$ value has no effect in the simulations and in the mean-field results.

Before  going on,  let us state one fundamental feature of the model that is independent if the update is synchronous or not. Suppose a focal player imitates a defection strategy. We state that any interaction of type (C,C) will never be replaced by (D,C). If some opponent adopts defection, the focal player must have at least one (C,D) or (D,D) interaction. But these interactions give  payoffs worst than (C,C). So (C,C) will never be replaced. This proves the existence of a lower bound for the fraction of cooperation given by the initial  fraction of mutual cooperation.  On the other hand, if the focal player imitates a D strategy, he/she will first seek for (C,D) interactions. If they are present, (C,D) will be replaced by (D,D). One can see that mutual cooperation is never destroyed and every exploitation is punished. 

For the usual game, where each player adopts a single strategy against all of its opponents, a single defector can invade a population of cooperators in infinity well-mixed population \cite{r1}.  The first remarkable feature of ``implicit punishment'' is that cooperation is evolutionary stable in well-mixed population for both synchronous and asynchronous update. If a mutant that adopts defection against everybody appears in a population where everybody is cooperating, the mutant initially earns a huge payoff. But as soon as others imitate it, the exploited (C,D) interactions will be replaced by (D,D), neutralizing the exploitations. What is simple, but remarkable, is that the interactions that are changing are just those in which the exploiter is involved and all of the other mutual cooperations are maintained. The ``implicit punishment'' will take place until the mutant cumulative payoff is equal to cooperator cumulative payoff. If we call $n_{exp}$ the quantity of defection adopted by the mutant exploiter,   the  payoff of the mutant exploiter and the  payoff of the cooperators are $P_{exp}=n_{exp}(b+\epsilon)$ and $P_{coop}=N-2$, respectively. By equating both expression we have the equilibrium fraction
\[
n_{exp}=\frac{N-2}{b+\epsilon}.
\]  

Let us now discuss the results obtained by numerical simulations with both synchronous and asynchronous updates. In the initial conditions, we set the players to cooperate with probability $0.5$ against each one of its opponents. This gives an initial cooperation fraction of $50\%$. For asynchronous update, cooperation cannot dominate  population but it can coexist with defection and assumes values near the lower-bound ($25\%$ for the initial condition assumed here). Fig. 1  shows simulation and mean-field approximation results for the asynchronous update.
For the synchronous update, cooperation dominates the entire population. Fig. 2 shows the simulation and mean-field approximation results for the synchronous update. One can see in Fig. 3  that, for short times, the synchronous update behaves like the asynchronous update. At the beginning, all of the exploitations are neutralized and only the initial mutual cooperation survives. After that, cooperation starts to increase very slowly until it dominates the population. So we can define a short-time regime and a long-time regime for synchronous dynamics.  The same qualitative result holds for large populations, but simulation time gets extremely huge for synchronous update. Fig. 4 shows a finite size analysis for the time spent to reach the minimum value of the cooperation fraction before cooperation dominates in the synchronous update. Note that as along as $N$ increases, $1/T$ goes to zero, implying 
that there is no long-time regime for $N\to\infty$. On the other hand, if we have $N$ large but finite, the long-time regime is the asymptotic regime, which is characterized by domination of cooperation.

\vspace{3cm}
\begin{figure}[h]
  \centering
  \centerline{\includegraphics[width=7.5cm,height=5.0cm]{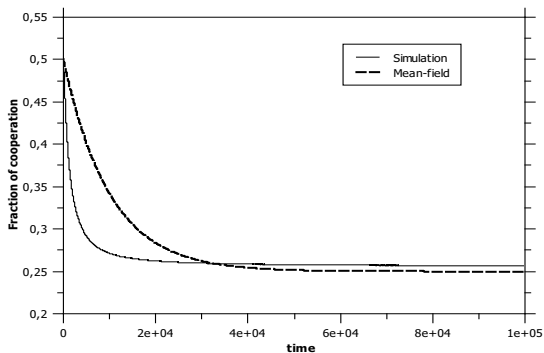}}
\caption{Fraction of cooperation for asynchronous update with $N=100$ and $b=2.0$. }\label{fig1}
\end{figure}

\begin{figure}[h]
  \centering
 \centerline{\includegraphics[width=7.5cm,height=5.0cm]{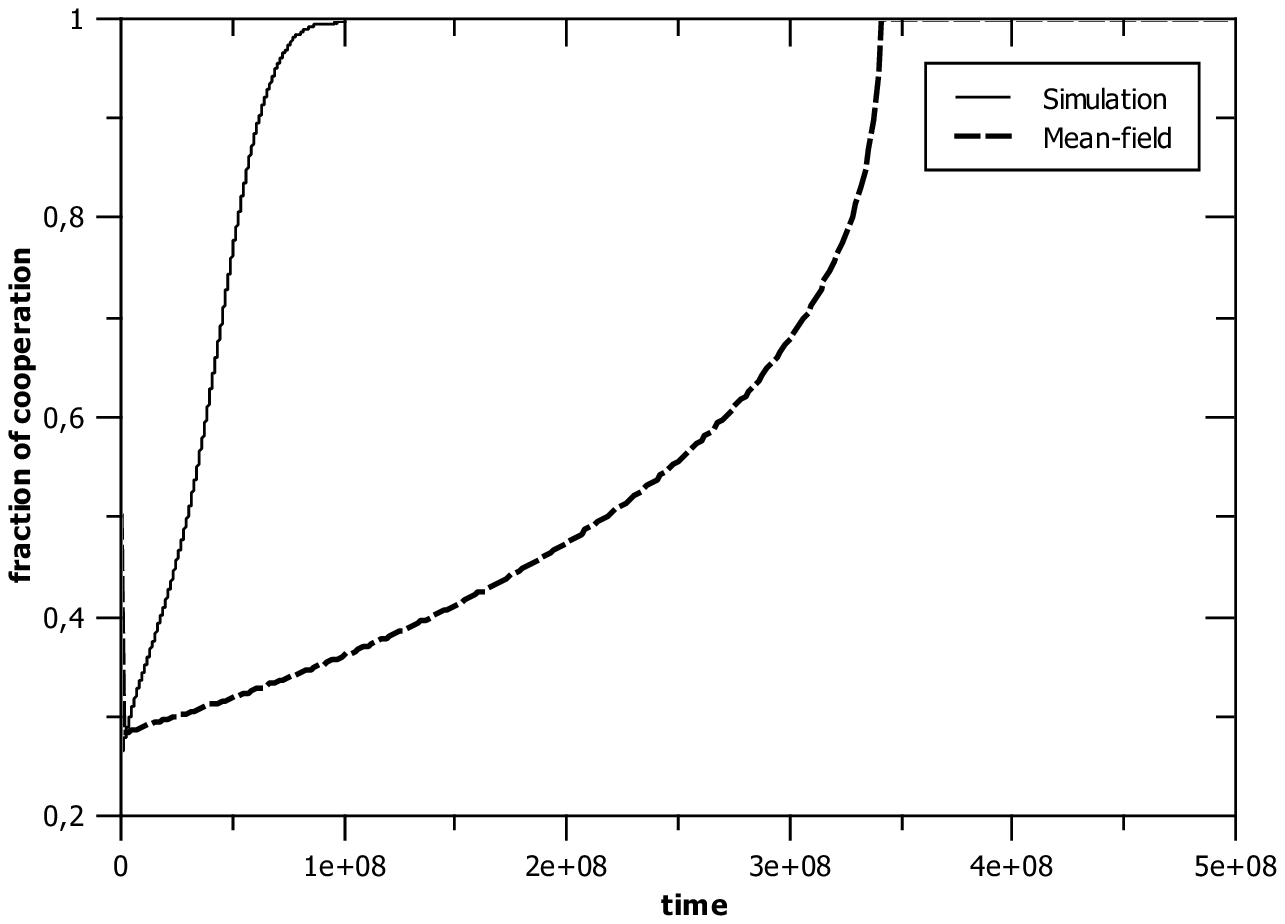}} 
\caption{Fraction of cooperation for synchronous update with $N=40$ and $b=2.0$. }\label{fig2}
\end{figure}

\begin{figure}[h]
  \centering
 \centerline{\includegraphics[width=7.5cm,height=5.0cm]{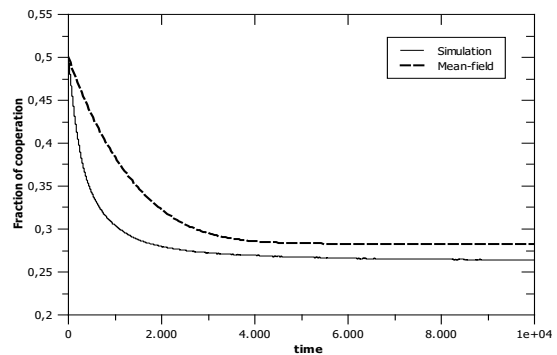}}   
\caption{Short-time regime of the cooperation fraction for synchronous update with $N=40$ and $b=2.0$.
 }\label{fig3}
\end{figure}

\begin{figure}
  \centering
 \centerline{\includegraphics[width=7.5cm,height=5.0cm]{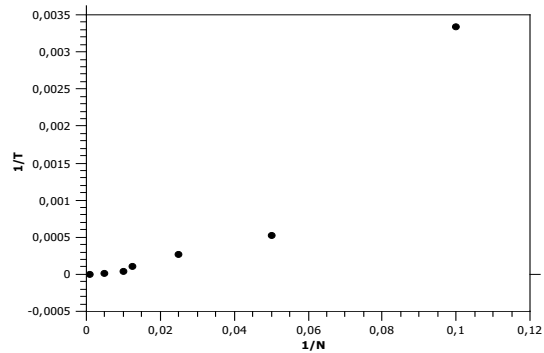}}   
\caption{The plot shows the relation $1/T \times 1/N$, where $N$ is the population size and $T$ is the time to reach the minimum value of cooperation for synchronous update.}\label{fig4}
\end{figure}

mean-field solution provides a good equilibrium analysis in well-mixed population if the usual game is considered \cite{r30}. But in structured population, it is not a good approximation \cite{r31}. Although in the present work we deal with well-mixed population, the nature of the ``implicit punishment'' model is not so simple. It is not obvious that a mean-field approach would work. So it is a remarkable result the fact that our mean-field approximation gives not only the stationary solutions, but fits reasonably the \textit{in silico} time evolution, although for the synchronous update it fits well only in the short-time regime. Let us now derive the mean-field solution. We first analyze the asynchronous  update followed by the synchronous one.

\begin{center}
  \textbf{mean-field approximation for the asynchronous update}
\end{center}

Let us first define a  local and a global interaction concentrations for a population of size $N+1$. If in an interaction a  player $i$ adopts  strategy A and its opponent adopts B, where $A,B \in\{C,D\}$,  we say that  player $i$ has a (A,B) interaction. Player $i$ can have $N_{AB}(i)$ (A,B) interactions. We define the local concentration of (A,B) as the fraction of (A,B) interactions around player $i$, namely $x_{AB}(i)=N_{AB}(i)/N$. For the  global concentration of (A,B) we define $x_{AB}=\sum_i N_{AB}(i)/((N+1)N)$.  

We first consider a typical player, that we call focal player. We are going to study the dynamics of the local concentration of (C,D), (D,D), (C,C), and (D,C) interactions around the focal player. Let $k_1$, $k_2$, $k_3$, and $k_4$ be the quantity of such interactions around the focal player. Note that $k_1+k_2+k_3+k_4= N$. Let us assume that the probability of having $k_1$, $k_2$, $k_3$, and $k_4$ interactions  are given by the respective global concentrations. The probability of a focal player configuration $(k_1,k_2,k_3)$ is given by
\begin{widetext}
\[
P_{k_1,k_2,k_3}=\frac{N!}{k_1!k_2!k_3!(N-k_1-k_2-k_3)!}x_{CD}^{k_1}x_{DD}^{k_2}x_{CC}^{k_3}x_{DC}^{N-k_1-k_2-k_3}
\]
\end{widetext}
We  consider the other nodes  as mean-field nodes. So the the local concentration of (C,D), (D,D), and (C,C) interactions around those nodes is  given by the configuration vector $N(x_{CD},x_{DD},x_{CC})$. Now we are going to derive the rate of variation of the local concentration around the focal player.

Let us derive the rate of increasing of the $(D,D)$ local concentration. Suppose that the focal player is on a $(k_1,k_2,k_3)$ configuration. Note that the payoff of this configuration of the focal player is
\[
PF=(N-k_1-k_2-k_3)b+k_3+k_2\epsilon.
\]   
There is just one transition that increases this quantity: (C,D) to (D,D). In this replacement, the focal player adopts C and imitates an opponent that adopts a D. So  at least one interaction of type (C,D) must be present. But the focal player can imitate the D strategy from (C,D) or (D,D) interactions. Let us focus on the first alternative, that happens with probability $k_1/N$.  The opponent associated with the (C,D) interaction has a payoff of
\[
PF_{CD}=b+(N-1)(x_{DC}b+x_{CC}+x_{DD}\epsilon).
\]
The probability of imitating the D strategy from (C,D) is given by
\[
\frac{[PF_{CD}-PF]\Theta(PF_{CD}-PF_k)}{bN},
\]
where $\Theta(x)=1$ if $x>0$, and $\Theta(x)=0$ otherwise.
The mean rate of increasing of (D,D) by one unit due to the imitation from (C,D) is  
\begin{widetext}
\[
W_{CD}^{+}=\sum_{k_1=1}^{N} \sum_{k_2=0}^{N-k_1}\sum_{k_3=0}^{N-k_1-k_2}\frac{k_1}{N}\frac{[PF_{CD}-PF]\Theta (PF_{CD}-PF)}{bN}P_{k_1,k_2,k_3}
\]  
Following the same lines we obtain the possibility of imitating from (D,D), that is given by
\[
W_{DD}^{+}=\sum_{k_1=1}^{N} \sum_{k_2=0}^{N-k_1}\sum_{k_3=0}^{N-k_1-k_2}\frac{k_2}{N}\frac{[PF_{DD}-PF]\Theta (PF_{DD}-PF)}{bN}P_{k_1,k_2,k_3}.
\]  
\end{widetext}
The same analysis can be done to calculate the rate of decreasing of $(D,D)$. There is just one transition involved, namely, (D,D) to (C,D). At least one (D,D) must be present. The C strategy can be imitated from (C,C) or (D,C). But note that the focal player cannot be currently adopting a (C,D), because in that case (C,D) would give the worst payoff. So imitating a C strategy would not change the quantity of (D,D). Following the previous steps, we can define
\[
P_{k_2,k_3}=\frac{N!}{k_2!k_3!(N-k_2-k_3)!}x_{DD}^{k_2}x_{CC}^{k_3}x_{DC}^{N-k_2-k_3},
\]
and we obtain that
\begin{widetext}
\[
W_{CC}^{-}=\sum_{k_2=1}^{N} \sum_{k_3=0}^{N-k_1}\frac{k_3}{N}\frac{[PF_{CC}-PF]\Theta(PF_{CC}-PF)}{bN}P_{k_2,k_3}(1-\delta_{k_3,0}\delta_{k_2,N}), \quad \mbox{and}
\]  
\[
W_{DC}^{-}=\sum_{k_2=1}^{N} \sum_{k_3=0}^{N-k_1}\frac{N-k_2-k_3}{N}\frac{[PF_{DC}-PF]\Theta(PF_{DC}-PF)}{bN}P_{k_2,k_3}(1-\delta_{k_3,0}\delta_{k_2,N}).
\]
\end{widetext}

 As all of the nodes have the same typical behavior, because the population is well-mixed, we can approximate the global concentrations by the local ones. The above expressions determine the rate of (D,D) variation by one unit. If we want the time derivative of $x_{DD}$, we need do divide the expressions by $N$ and multiply by a factor of two, because there is the contribution of the opponents update . So we have
\[
\frac{dx_{DD}}{dt}=2\frac{1}{N}(W_{CD}^{+}+W_{DD}^{+}-W_{CC}^{-}-W_{DC}^{-}).
\]
Following the same reasoning, one can see that $x_{CC}$ does not change in time. Finally, as all of the mean-field variables are normalized to one, we obtain that 
\[
\frac{dx_{CD}}{dt}=-\frac{1}{2}(\frac{dx_{DD}}{dt}+\frac{dx_{CC}}{dt}).
\]

We can simplify further these expressions if we replace $k_1$, $k_2$ and $k_3$  inside the payoff expressions of the focal player by the expected value of such quantities given by the configuration probability: $Nx_{CD}$, $Nx_{DD}$, and $Nx_{CC}$, respectively. With this extra approximation the $\Theta$ function can be easily evaluated in the limit of large $N$ and we have the following equation: 
\begin{eqnarray*}
\frac{dx_{DD}}{dt}&=&2\frac{1}{N^2}\{ x_{CD}+x_{DD}\frac{\epsilon}{b}[1-(1-x_{CD})^{N-1}] \\
&-&\frac{x_{CC}}{b}[(1-x_{CD})^{N-1}-(x_{CC}+x_{DC})^{N-1}]\} 
\end{eqnarray*}

This equation can be solved numerically. Fig.1 shows the numerical solution and the simulation results. Note that this approximation furnishes good results when compared with \textit{in silico} evolution. For the initial condition used here, we have that the terms inside the parentheses are $0.75$ powered to $N-1$ and $0.5$ powered to $N-1$. If $N$ is large, the terms that are powered to $N$ are very small and they can be neglected, at least for short times. This gives the following simplified equations,
\[
\frac{dx_{DD}}{dt}=2\frac{1}{N^2}(x_{CD}+\frac{\epsilon}{b}x_{DD}), \quad \mbox{and} \quad \frac{dx_{CC}}{dt}=0.
\]
The solution of these equation are straightforward, 
\[
x_{DD}=x_{DD}^0+ \frac{\frac{\epsilon}{b}x_{DD}^0+x_{CD}^0}{\frac{1}{2}+\frac{\epsilon}{b}}[1-\exp(-\frac{2}{N^2}(\frac{1}{2}+\frac{\epsilon}{b})t)],\quad \mbox{and}
\]
\[
x_{CC}=x_{CC}^0,
\]
where the index $0$ refers to the initial conditions. If we set $\epsilon=0$, just for simplicity, the solution reaches the fixed point 
\[
x_{DD}^{\infty}=x_{DD}^0+2x_{CD}^0, \quad x_{CC}^{\infty}=x_{CC}^0,  
\]
\[
\quad \mbox{and} \quad x_{CD}^{\infty}=x_{DC}^{\infty}=0.
\]
One can see that only the initial mutual cooperation can be maintained and all of the other interactions are mutual defections. Note that all of the exploitations are neutralized. Note that this approximation gives good results if compared to simulation data.

\begin{center}
  \textbf{mean-field approximation for the synchronous update}
\end{center}

Let us treat the synchronous case. Now (C,C) can increase, because  it is possible to have a  (D,D) to (C,C) transition whenever two players make a (D,D) to (C,D) transition in their shared (D,D) interaction. This is an essential feature of the synchronous model. This kind of transition does not take place in asynchronous update and that is the reason why cooperation assumes the lower bound value in the asynchronous case.  We can approximate the rate of this transition by 
\[
\frac{dx_{CC}}{dt}=\frac{1}{Nx_{DD}}(W_{CC}^{-}+W_{DC}^{-})^2.
\]
Let us explain the term in the denominator. If the focal player
makes a (D,D) to (C,D) transition on an specific interaction, the mean-field player associated to this specific interaction should choose exactly this interaction, what happens with probability $1/Nx_{DD}$. If we perform  the same simplifications that was already done for the asynchronous case, we have
\[
\frac{dx_{CC}}{dt}= 2\frac{1}{N^4}\{\frac{x_{CC}}{b}[(1-x_{CD})^{N-1}-(x_{CC}+x_{DC})^{N-1}]\}^2, 
\]
\begin{eqnarray*}
\frac{dx_{DD}}{dt}&=&2\frac{1}{N^2}\{ x_{CD}+x_{DD}\frac{\epsilon}{b}[1-(1-x_{CD})^{N-1}]\\
&-&\frac{x_{CC}}{b}[(1-x_{CD})^{N-1}-(x_{CC}+x_{DC})^{N-1}]\}
\end{eqnarray*}

Fig. 2  and Fig. 3  show the numerical solution of these equations. One can see from the above equations that $x_{CC}$ increases much slower than $x_{DD}$. For the initial condition assumed here, $x_{CC}$ time derivative at the beginning is almost zero, because the values inside the brackets are equal to $0.5$ powered to $N$. But when  evolution starts,  great part of (C,D) interactions are changed to (D,D) and  $x_{CD}$ is reduced to some value near to zero making  $x_{CC}$ to increase faster. So we have two regimes: short-time regime, when $x_{CC}$ is kept almost constant around its initial values, and long-time regime, when $x_{CC}$ starts to increase faster. Fig. 2  and Fig. 3 show both cooperation evolution regimes. For short times, if we discard the terms that are powered  to $N-1$, we have the same solution as  the asynchronous case. This means that cooperation assumes a value near its lower bound value, given by the initial mutual cooperations. But for long-time regime, $x_{CD}$ is near zero and $\dot{x}_{CC}$ cannot be neglected. So $x_{CC}$ starts to increase until it become equal to one. So for sufficient long times,  the stationary solution is
\[
x_{CC}^{\infty}=1 \quad \mbox{and} \quad x_{CD}^{\infty}=x_{DC}^{\infty}=x_{DD}^{\infty}=0.
\]
 Note that for the short-time regime, shown in Fig. 3, the mean-field approximation fits well when compared to \text{in silico} evolution. For the long-time regime, the time evolution of the mean-field solution does not fit well, although it gives the right stationary solution.

From the above expressions and simulation data, we see that the lower value of cooperation is reached very fast in the synchronous update. But as long as population size gets bigger, this value is reached very slowly (see Fig. 4). Besides that, if $N$ is large, $x_{CC}$ increases very slowly. By these reasons, for large $N$, in the synchronous update the population seems to be wrapped in the lower value of cooperation, although what is actually happening is that cooperation is slowly increasing, spreading until dominates the entire population.  

\begin{center}
  \textbf{Conclusion}
\end{center}

Here we analyzed the model that allows the individuals to choose different strategies against different opponents in well-mixed populations for both synchronous and asynchronous update. First we showed that cooperation is evolutionary stable for both synchronous and asynchronous update, what means that a defector mutant cannot invade a population of cooperators. We also showed,  for a initial condition of $50\%$ of cooperation, that for synchronous update cooperation always dominates while for asynchronous update the cooperation fraction assumes the lower bound given by the initial mutual cooperation. For the synchronous update, population dynamics exhibits a short-time behavior that is similar to the asynchronous update, but for suficient long times, cooperation spreads for large $N$ but finite. The crucial difference between synchronous and asynchronous is that in synchronous update it is possible to have a simultaneous update that allows a (D,D) to (C,C) transition. This does not happen in asynchronous update. In a preview work, the same model was analyse in a square lattice with synchronous update. Here we showed that the synchronous update is crucial for cooperation dominance while network reciprocity effects are not so important. Although the asynchronous update does not provide cooperation dominace, it allows cooperation to survive at its lower bound value. Note that this result in asynchronous update is still better than those of the usual game.

\acknowledgments
The authors  thank to CNPq and FAPEMIG, Brazilian agencies.

\smallskip


\noindent {\small $^\ast$To whom correspondence should be addressed (e-mail: wardil@fisica.ufmg.br).}

\vfill

\begin{thebibliography}{99}

\bibitem{r0} J. Weibull
                 { \sl Evolutionary game theory} ,
                  MIT Press, Cambridge, USA (1995).


\bibitem{r1} M. A. Nowak,
                 { \sl Evolutionary Dynamics: Exploring the Equations of Life} ,
                  1rd ed., Harvard University Press (2007).


\bibitem{r23} M. A. Nowak
             Science {\bf 314} 1560-1563 {2006}.

\bibitem{r3}  J. Henrich, N. Henrich
             Cognitive Systems Research  {\bf 7}, 220-245 (2006).
             
\bibitem{r11}  D. Rand, H. Ohtsuki, M. A. Nowak
             J. Theor. Biol.  {\bf 256}, 45-57 (2009).


\bibitem{r10}  B. Rockenbah, M. Milinski
             Nature  {\bf 444}, 718-723 (2006).



\bibitem{r12}  H. Ohtsuki, Y. Iwasa, M. A. Nowak
             Nature  {\bf 457}, 79-82 (2009).

             
             
\bibitem{r13}  A. Dreber, D. G. Rand, D. Fudenberg, M. A. Nowak
             Nature  {\bf 452 }, 348-351 (2008).
             
\bibitem{r19}  S. Assenza, J. Gomez-Gardenes, and V. Latora
             Physical Review E  {\bf 78 }, 017101-1 (2008).
             
\bibitem{r20}  J. Gomes-Gardenes, M. Campillo, L. M. Floria, and Y. Moreno
             Physical Review Letters {\bf 98 }, 108103-1 (2007).

\bibitem{r21}  Z. Rong, X. Li, and X. Wang
             Physical Review E {\bf 76 }, 027101-1 (2007).

\bibitem{r22}  J. Poncela, J. Gomez-Gardenes, L. M. Floria, and Y. Moreno
             New Journal of Physics  {\bf 9 }, 184 (2007).
             
\bibitem{r27} M. Nakamaru, H. Matsuda, and Y. Iwasa 
             J. Theor. Biol. {\bf 184}, 65-81 {1997}.

\bibitem{r32} E. A.Sicardi, H. Fort, M. H.Vainstein, and J. J.Arenzon
             J. Theor. Biol.  {\bf 256}, 240-246 (2009).
             

\bibitem{r6}  E. Feher, S. Gachter
             Nature  {\bf 415}, 137-140 (2002).

\bibitem{r8}
            R. Boyd, H. Gintis, S. Bowles, P. J. Richerson
            P. Nat. Acad. Sci. USA {\bf 100}, 3531-3535 (2003).

\bibitem{r9}  J. Henrich, R. Boyd
             J. Theor. Biol.  {\bf 208}, 79-89 (2001).


\bibitem{r15}  L. Wardil, J. K. L. da Silva
             EPL  {\bf 63}, 38001 (2009).
             
\bibitem{r29} H. J. Blok, and B. Bergersen 
             PNAS {\bf 90}, 7716-7718 {1993}.       

\bibitem{r28} H. J. Blok, and B. Bergersen 
             Physical Review W {\bf 59}, 3876-3879 {1999}.             
 

\bibitem{r18}  F.C. Santos and J. M. Pacheco
             Physical Rewiew Letters {\bf 95}, 098104-2 (2005).
             
\bibitem{r30} G. Szabó, and G. Fathb
            Physics Reports {\bf 446}, 97-216 (2007).


\bibitem{r31} G. Szabó, J. Vukov, and A. Szolnoki
             Phys. Rev. E {\bf 72}, 047107 (2005).



           
 

\end{thebibliography}
\end{document}